\documentclass[fleqn,10pt]{wlscirep}

\usepackage{hyperref}
\graphicspath{{./}}

\newcommand{\p}[1]{\mathbf{#1}}

\title{
\emph{Active} and \emph{reactive} behaviour in human mobility: the influence of attraction points on pedestrians} 

\author[1,+,*]{Mario Guti\'errez-Roig}
\author[1,+,*]{Oleguer Sagarra}
\author[2]{Aitana Oltra}
\author[3]{John R.B. Palmer}
\author[2,3]{Frederic Bartumeus}
\author[1]{Albert D\'iaz-Guilera}
\author[1]{Josep Perell\'o}

\affil[1]{Departament de F\'{\i}sica Fonamental, Universitat de Barcelona. Mart\'i i Franqu\'es 1, E-08028 Barcelona, Spain}
\affil[2]{Centre d'Estudis Avan\c cats de Blanes (CEAB), CSIC, Acc\'es a la Cala Sant Francesc, 17300 Blanes (Girona), Spain}
\affil[3]{Centre de Recerca Ecol\`ogica i Aplicacions Forestals (CREAF), Campus de Bellaterra (UAB) Edifici C, 08193 Cerdanyola del Vall\`es, Spain}
\affil[*]{mariogutierrezroig@ub.edu, osagarra@ub.edu}
\affil[+]{these authors contributed equally to this work}

\begin{abstract}
Human mobility is becoming an accessible field of study thanks to the progress and availability of tracking technologies as a common feature of smart phones. We describe an example of a scalable experiment exploiting these circumstances at a public, outdoor fair in Barcelona (Spain). Participants were tracked while wandering through an open space with activity stands attracting their attention. We develop a general modeling framework based on Langevin Dynamics, which allows us to test the influence of two distinct types of ingredients on mobility: \emph{reactive} or context-dependent factors, modelled by means of a force field generated by attraction points in a given spatial configuration, and \emph{active} or inherent factors, modelled from intrinsic movement patterns of the subjects. 
The additive and constructive framework model accounts for the observed features. Starting with the simplest model (purely random walkers) as a reference, we progressively introduce different ingredients such as persistence, memory, and perceptual landscape, aiming to untangle \emph{active} and \emph{reactive} contributions and quantify their respective relevance. The proposed approach may help in anticipating the spatial distribution of citizens in alternative scenarios and in improving the design of public events based on a facts-based approach.
\end{abstract}

\begin{document}
\flushbottom
\maketitle
\thispagestyle{empty}

The development of information and communications technologies (ICT) is changing the way we interact with the world, and by extension the way we perform science. We have entered the big data era \cite{King2011} which, in principle, puts in the hands of researchers enormous possibilities to monitor, study and understand human activities. Among such technologies, the development of readily available, cheap and reliable geo-localized devices has fostered the field of studies in human mobility. Computer vision analysis \cite{Karamouzas2014}, Bluetooth devices \cite{Bollo2005,Yoshimura2012}, RFID signal intensity \cite{Barrat2010,Stehle2011}, mobile phone calls \cite{Gonzalez2008,Palchykov2014,Calabrese2011} and geo-localized social network feeds \cite{Noulas2012} represent just a handful of the methods and data sources that have now been successfully exploited for this purpose.

Despite the recent technological advances, two important challenges remain for the scientific community: accessibility and data biases. On one hand, data openly available for human mobility research is scarce and somewhat restricted. ICT companies are reluctant to freely share their data with the scientific community, and when they do so it is often with strict constraints that pose challenges for the reproducibility of experiments. On the other hand, human mobility data gathered directly from traditional research volunteers is often limited by privacy considerations and scalability. In the present paper we propose that the emerging citizen science research model \cite{Hand2010,irving,Sagarra2015} offers an alternative strategy that can solve the problems of data quality, control, property rights, accessibility, and privacy, offering reproducible results. We use this strategy to investigate human mobility at an outdoor science fair in Barcelona (Spain). By gathering location information in a crowded public space and making volunteers active participants in the scientific process, we are able to easily construct a unique high-resolution and open-access data set.

In the context of human mobility, cell phone GPS receivers offer a good technological option for gathering location information \cite{Palmer2013}, one that also bridges the field of human movement in open spaces with that of behavioural ecology \cite{Gonzalez2008,Brockmann2006,Raichlen2014}, with its rich and consistent literature on home-ranging and foraging. Although the resolution of GPS-based location estimates often drops in urban areas (e.g., due to building shadowing) GPS nonetheless offers the ability to track individuals at mid-range spatial scales (e.g. hundreds of meters to several kilometers) and may be augmented by location information from cellular networks, wifi routers and other sources. Moreover, knowledge extracted even at mid-range scales can be critical to understanding and improving the quality of life in cities, as it closely manifests the relationship between people and the urban environment.

A key aspect to understanding organisms' movement patterns, including those of humans, is determining how {\it active} and {\it reactive} behavioural components are exploited depending on scales and information availability \cite{Bell1991,Dusenbery1992,Mischiati2015}. Depending on the state of the organism (e.g. level of hunger, stress) and the amount and quality of the available information, movement is tightly capitalized by sensory or memory information (\emph{reactive} motion) or else by a more explorative inherent component (\emph{active} motion). Importantly, exploratory movement may also be guided by sensors and cognitive processing (past experiences), but the motor connections between cause and effect should be considered less explicit and time-delayed. \emph{Reactive} movement is then more promptly associated to movement driven by external triggers, whereas \emph{active} motion is more likely context-independent and internally driven\cite{Bell1991,Dusenbery1992,Mischiati2015}. One may also connect the binary \emph{reactive}/\emph{active} distinction to the classic debate in movement ecology on whether external or internal factors govern movement \cite{Nathan2008}. In any case, the distinction between \emph{reactive} and \emph{active} motion should be taken cautiously and in relative terms, with \emph{active} motion viewed as involving more free movement actions than \emph{reactive} motion, but not being the exclusive domain of such actions.

Here we model and characterize human movement at the Barcelona science fair in terms of \emph{active} and \emph{reactive} components. In our experiment, participants were invited to wander around an open space containing the fair's activity stands while tracking themselves using a mobile phone application. Individuals could freely move around and access any of the stands. The space in which they moved was not isotropic---it included obstacles, paths trails, and forbidden walking areas---but the stands were set up to clearly attract most of the attention and drive movement dynamics in this ludic, open-air event. 

We propose an additive framework model to characterize participants' motion, using a complex systems science approach to account for some of the observed features. Starting with the most simple model (purely random walkers) as a reference, we progressively introduce different ingredients such as persistence, memory, and perceptual landscape, aiming to untangle \emph{active} and \emph{reactive} contributions and quantify their respective relevance. We then compare the limitations and strengths of the proposed models, and discuss the delicate balance between complexity and accuracy when modeling human movement. The framework we propose is flexible enough and sufficiently easy to handle to be used as a tool to better understand mobility in other contexts.

\section*{Results}
\subsection*{The experiment, the data and some basic definitions}
\label{sec_exp}

We carried out the experiment, called Bee-Path (BP), at Barcelona's annual science festival, a major event held in an public park and promoted by the city council (see Methods for further details). During two consecutive days, participants were tracked as they wandered around the park to visit fair stands that offered a variety of activities. While the participants had access to some information about the activities held in the stands (and possibly some prior knowledge about the environment), they were recruited upon entering the space, as shown in Fig. \ref{fig1}, and hence are assumed to have had no previous direct experience in exploring the fair.

After removing outliers from the data (see Supplementary Information, SI), we ended up with 4,994 GPS locations, corresponding to 27 participants and roughly 27 hours of experiment. The timestamped GPS locations for each subject were converted into stop-and-run individual walks using a two-step procedure (see Methods and Fig. \ref{fig1}): First, locations were flagged as either \textit{moving} or \textit{stopped}. Second, successive locations flagged as \textit{stopped} were used to determine stop time lengths, and the successive locations flagged as \textit{moving} were grouped into flights using the so-called rectangular grid criteria \cite{Rhee2007} (see Fig. \ref{fig1} for further details). Each stop has a time duration $\Delta t_s$ and a location $\p{x}_s$ defined as the average position of its stopped points. Conversely, each flight has a time duration $\Delta t_f$ and a length $\Delta r_f$ defined as Euclidean distance between starting and ending points. These two magnitudes also allow us to obtain an average velocity for each flight $v = \Delta r_f/\Delta t_f$. Each participant has a track defined by these flights and stops, and each track has an associated mean velocity, calculated as the average of its flight velocities. There are a total of 350 flights and 300 stops.

Although the participants vary in demographic characteristics, we find no particular profile in terms of age or gender (see discussion in SI) and we assume population homogeneity, pooling all data to improve statistical discernment between \emph{stopping} and \emph{moving} states, better detect main orientation features, and minimize the effect of location errors caused by GPS noise. Also crucial is the assumption that the stop-and-run tracks obtained adequately characterize key aspects of the actual movement, which may be intrinsically assumed to be a continuous process in space and time. The most we can do is to consider and evaluate potential biases in the conversion from discrete GPS locations to stop-and-run walks by exploring how parameter choices in the movement processing algorithm affect results (see SI where also the biases on an isotropic Random Walk are analytically evaluated).
 
Considering these preliminaries, we analyse movement features from a two-fold perspective. We first assess participants' overall space use by computing the population-level utilitisation distribution. Second, we take the stop-and-run walks approach to compute movement metrics that help us to understand key features of movement as well as to evaluate how the empirical data fits into different classes of proposed models.

\subsection*{Movement patterns}
\label{mov}

We compute the population-level utilisation distribution in two-dimensional space using a dynamic Brownian Bridge Movement Model (dBBMM) \cite{Kranstauber12, Bullard99,HorneEtal07, NelsonEtal11}, which assumes observed locations are bridged by Brownian motion, with the diffusion coefficient adjusted for each bridge based on observed variance in the trajectory. We apply this model to all tracks to calculate, for each cell in discretized observation space, the relative frequency of participants' presence during the experiment (see Methods for further details). Top left Fig. \ref{fig2} shows that participants were mainly concentrated around the stands, albeit with this concentration seeming to decrease as we move further away from the starting point of the experiment. Therefore, the specific configuration of the stands, placed along the x-axis, has a strong effect on the mobility of the participants. The space shows the non-isotropic character of urban areas and spaces, which in many cases is purposeful (stand distribution and accessibility were deeply discussed by the fair's organizing team). The polar plots in top right Fig. \ref{fig2} further confirm this finding: The flight orientation is biased in the direction of the promenade --the x-axis-- where the concentration of stands is high, while the asymmetry between the positive, $0º$, and negative direction, $180º$, in the x-axis reveals the general tendency of the people to form a flow from the entrance of the park, at the left of the map, to the last stand of the fair, at the right. 

The empirical stopping time statistics can be observed in bottom left Fig. \ref{fig2}. The observed stopping time Complementary Distribution Function (CDF) can thus be well fitted by a weighted double exponential law with two different characteristic decaying times,
\begin{align}
P(\Delta t_s > t) = \omega \exp({-t/\tau_1}) + (1-\omega)\exp({-t/\tau_2}).
\label{Eq1}
\end{align}
The shortest characteristic decay time is around $30$ seconds and corresponds to reorientation processes. As shown in Fig. \ref{fig3}, the shortest stops (smaller circumferences) appear to be randomly distributed over the fair space, indicating that this timescale may be related to orientational process not directly driven by the fair stands (\emph{reactive} behaviour) but linked instead to other factors. In contrast, Fig. \ref{fig3} also shows that the longest stops are much closer to fair stands (spots in yellow) where activities are happening. The characteristic scale of these flights is around 10 minutes (largest characteristic scale in the CDF of Eq. (\ref{Eq1})). The stop lengths are much smaller than the time span of the activities held in the stands, suggesting limited attention capacity \cite{Chee2007} by the fair participants. The fair included science outreach activities, such as talks and hands-on workshops, with durations generally between $30$ minutes and $1$ hour or in some cases even longer. Some stands hosted activities with no specific duration, lasting an entire morning or afternoon, or even the whole day. Based on the divergence between the time length of the stops at the stands and the length of the activities prepared for the fair, we have advised the organisers to aim for shorter, fresher and more dynamic activities to better capture audience attention in future editions of the fair.

Figure \ref{fig2} also shows that the flight length CDF displays a single exponential decay
\begin{align}
P(\Delta r_f > r) = \exp\left[{-(r - R_{stop})/\lambda}\right] \qquad \text{if}\quad r>R_{stop}.
\end{align}
We note that the distribution has a natural lower bound when $r=R_{stop}$ forced by the rectangular grid criteria for the flights algorithm described in the bottom panel of Fig. \ref{fig1}. Due to the relatively small park dimensions and moderate densities, one could neither expect very large displacements nor identify heavy-tailed flight behaviour \cite{Gonzalez2008}. Results show however a relatively slow decay compared to the overall scale of the fair space ($330\times 145$ meters), with a characteristic length of $\lambda=27$m. This scale is large
enough to consider the presence of such long flights as unexpected, given the moderate densities (non-crowded) and the small and bounded environment in which fair activities were taking place. Our experimental data shows that some participants skipped stands, sometimes making larger flights more probable, and exhibiting the type of short-range directional persistence commonly found in animal movement \cite{Turchin1998}. An extended discussion is provided in the SI.

A final feature worth studying is the distribution of flight
velocities. This is highly peaked around $0.6\pm0.3$m/s which is below standard human velocity speed (see SI). The small average flight velocity detected is probably the result of a moderate density of attendants to the fair in combination with the multiple activities available to visitors. Moreover, such a reduced variance in flight velocity also suggests low individual variability and the presence of an homogeneously behaved population. We computed velocity distributions discarding the stops, only using the moves or flight segments. 

\subsection*{The Bee-Path attractivity model framework}\label{sec_mod} 

Pollinators, such as bees, move across the landscape looking for resources (within flowers) by following a stop-and-run process\cite{Bell1991,Turchin1998}. The resulting stop-and-run paths emerge as a combination of \emph{reactive} (extrinsic) and \emph{active} (intrinsic) individual behaviour. Similarly, the human movement observed at the fair can be modelled as a stop-and-run process driven by two main components: external inputs and intrinsic motion. Individuals largely wandered through the fair according to their instincts or some vague expectations (context independent decisions) but, from time to time, they could \textit{feel the attraction} of specific stands hosting specific activities. We develop here minimal models of increasing complexity (BP model framework) that may help us to distinguish external from internal drivers of motion, and define the key features of the movement observed within the fair.

To describe the position $\p{r}(t)$ of an individual at time $t$, we use a two-dimensional stochastic model based on Langevin Dynamics \cite{Gardiner}, which describe the trajectory of a particle moving randomly due to a thermal bath (\textit{internal energy}) and subject to a field of forces (\textit{external energy}). We assume individuals have no inertia: Although humans clearly have memory of past events, our assumption here is that their orientation decisions during the fair are mostly taken with respect to actual circumstances \cite{Codling2008}. The dynamical equation for the BP model framework, discretized in intervals of $\Delta t$ and in the It\^o sense \cite{Gardiner} then reads, 
\begin{equation}
\p{v}(t)=\frac{\p{r}(t+\Delta t)-\p{r}(t)}{\Delta t} = \dfrac{1}{\gamma}\p{\nabla }V(\p{r}(t)) + \sqrt{\frac{2}{\gamma\beta \Delta t}} \rho \p{\hat{u}}(\theta(\kappa,\Delta t))= \dfrac{1}{\gamma} \left(\p{F_R} + \p{F_A} \right), \qquad \p{\hat{u}}(\theta) = (\cos\theta, \sin \theta);
\label{eq:langevin}
\end{equation}
where the first term stands for the potential landscape capturing the field of forces (\textit{reactive} component, $\p{F_R}$) while the second refers to the random fluctuations on the movement (\textit{active} component, $\p{F_A}$). The potential landscape $V(\p{r})$ is generated by the joint effect of fourteen different sources of identical shape, which are located in the same positions as the stands in the experimental setup (see Fig. \ref{fig1}).

We have thus dynamics governed by two contributions, the \textit{reactive} one, which depends on the attraction form of the external inputs and the \textit{active} one, whose intensity is controlled by the parameters $\beta$ (random unit radial contribution) and $\kappa$ (random angular contribution) which simulate the inherent movement stochasticity \cite{Codling2008}. Drag parameter $\gamma$ is manually set to unity following the assumption that all individuals are equally susceptible to the action of both the \emph{reactive} and \emph{active} forces (a change of this parameter would only redefine the timescale of the process). Concerning the form of the potential for the attraction wells, a single stand centred at $\p{r}_0$ is represented as a gravitational-like attractive potential well generated by a non punctual spherical attractive mass\cite{Gardiner} 
\begin{equation}
\label{eq:V2}
V(\p{r}|\p{r}_0)=
\left \{
\begin{array}{l l}
-V_0\frac{\sigma}{|\p{r-r_0}|} & \text{if } |\p{r-r_0}|\geq\sigma\\
- \frac{3}{2}V_0 + \frac{1}{2}V_0\frac{|\p{r-r_0}|^2}{\sigma^2} & \text{if } |\p{r-r_0}|<\sigma .
\end{array}
\right.
\end{equation}
The two constants of the potential are fixed in terms of the radius of the non-punctual spherical region $\sigma$ (related to the size of the stands) and in terms of the intensity of the potential outside this circular region $V_0$ (related to the attraction force). Intensity solely depends on the Euclidean distance $|\p{r-r_0}|$. We therefore construct a set of attractive wells located at each of the stands of the fair as shown in Fig. \ref{fig3}. For simplicity, both $\sigma$ and $V_0$ are taken as equal for all stands, with the intensity of the stands $V_0$ adjusted from real data, and $\sigma=4$m chosen according to the actual physical area of the original stands.

We consider an individual to be trapped by an attractor when she approaches a given stand for the first time and comes within a distance shorter than $\sigma$; the other potentials then deactivate, representing that her attention is fully focussed on the activity of that stand. Under this scenario, we observe that the \emph{stopping} locations of the motion dynamics can be described by the classic Kramers problem \cite{Risken} extended to two dimensions, which evaluates the statistics of the trapping times. The classic exponential decay distribution of the Kramers problem is consistent with empirical \emph{stopping} time distribution (longest stops) as shown in Fig. \ref{fig2}. The ratio of parameters $V_0$ and $\beta$ determines the decaying ratio in the distribution of stops (see SI for a complete exploration of the parameters space). To incorporate memory effects in our modelling approach, we consider that after an individual has visited one of the attractive wells (a stand) and leaves it (being further than distance $\sigma$), the subject ceases to feel its attraction and henceforth the attractive well is deactivated for this walker. 

Therefore, the motion in our experiment is described by Eq. (\ref{eq:langevin}) with the potential $V(\mathbf{r}(t))$ taking the explicit forms in Eq. (\ref{eq:V2}) depending on the position of the subjects. They will be most of the time in a \emph{moving} state described by the first expression corresponding to $|\mathbf{r}-\mathbf{r_0}|\geq \sigma$. Occasionally subjects are \emph{trapped} inside a potential well described by the second expression. Therefore, dynamics are analogous to the Kramers problem \cite{Risken} and our processing algorithm labels them in a \emph{stopping} state. We thus see how this description bridges \emph{stop-and-run process} from behavioural ecology to the continuous in time BP model framework. 

\subsection*{Shaping the \emph{active} and the \emph{reactive} passive components.} 

The BP model framework proposed in Eq. (\ref{eq:langevin}) is highly flexible and makes it possible to clearly identify the basic ingredients of movement in the fair, recovering some of the empirical features of the experimental data. The used model contains a maximum of four free parameters ($\kappa$, $\beta$, $V_0$, and $\sigma$) although $\sigma$ is defined by the geometry of the attraction wells. We thus fix $\sigma=4$ m.

We can disentangle different aspects by playing with free parameters of the BP model framework and then simulate the subsequent four possible types of dynamics. Firstly, we can consider a pure random walk (RW) without attraction wells so that individuals have only an \emph{active} component ($V_0=0$, $\kappa=0$ and $\beta$ remaining as the only free parameter). Secondly, we can test another purely \emph{active} scenario where the only new ingredient is a persistence on keeping the direction of each previous step with a non-isotropic noise ($V_0=0$ while $\kappa$ and $\beta$ are free parameters). This case is known as a Correlated Random Walk (CRW) in the literature \cite{Codling2008}. Thirdly, we can observe a different situation where the individuals \emph{feel} the attraction wells but the underlying movement is a random walk such that \emph{active} and \emph{reactive} components are present ($\kappa=0$ while $V_0$ and $\beta$ are free parameters). We call this case Potential-driven Random Walk (PRW). Finally, we can combine both persistence and potential landscape ($\kappa$, $V_0$, $\beta$ are free parameters) in what we call Correlated Potential-driven Random Walk (CPRW). We have selected the values of the parameters for the PRW ($V_0$ and $\beta$) and CRW ($V_0$, $\beta$, and $\kappa$) models by optimizing their performance when looking at the empirical stop statistics as carefully discussed in the SI. The RW and CRW models take the same values as those from the equivalent models with potential attractive wells (PRW and CPRW).

The presented framework aims to study and identify the minimum ingredients necessary to capture the essential traits of human mobility at the scales studied. By means of hypothesis testing using the different proposed dynamics, we are able to assess the effect each additional parameter has on the observed movement patterns. The resulting statistics for each scenario are shown in Fig. \ref{fig2}. Both the the RW and CRW cases are obviously taken as null models. Lacking spatial coherence, they are unable to reproduce the spatial usage and orientation of the subjects or their stopping statistics. It is noteworthy, however, that the RW dynamics generate a flight velocity distribution (that of a Maxwell-Boltzmann distribution, see SI) close to the one observed in empirical data, indicating that a random noise term is a necessary addition to the model. As expected, the addition of correlation in the movement direction (CRW) leads to long flights (with an almost constant velocity), yet very large values of correlation are needed in order to partially reproduce the tail of the flight distribution (and the obtained results are not smooth due to the fact that all flights generated are excessively rectilinear, and this generates bias on the movement processing algorithm, see SI). 

Adding the presence of potential wells significantly increases the accuracy of the model, validating the hypothesis of perceptual landscapes and stressing the need of including \emph{reactive} components. Both the spatial distribution and temporal statistics of stops are recovered (with two distinctive time-scales), and the flight's velocities are again reproduced. The flight orientation is also distinctively shifted towards the direction of the fair promenade. The presence of long flights and the distribution of flight orientation, however, cannot be solely explained by the presence of attraction wells, since the absence of correlation in the noise term severely limits the probability of long displacements, which as discussed earlier is a distinctive feature of the observed data. Long flights cannot be accounted for by the dynamics presented, and are probably a result of a combination of interaction effects between individuals and the crowd, and directional bias due to the existence of a preferent direction imposed by the fair's main promenade. Confirmation of this fact is given by the CPRW dynamics, which include attraction and correlation effects, yet are still unable to capture the abnormally long (but not heavy-tailed) detected flights. Moderate improvements are detected on flight orientation and length distribution, yet the short scale of the stop statistics is lost. Even adding an extra ingredient accounting for selection of destinations by each individual -- a probability for each potential well not to be considered by the individual -- worsens the predictive power of the model (see SI).

Our approach links the two different contributions (\textit{active} and \textit{reactive} component) to the dynamics in a straightforward way. Furthermore, our approach through Eq. (\ref{eq:langevin}) makes it possible to quantify the balance between the competing components as expressed by the relative contribution to power developed by each of the acting forces $\p{F}_R$ and $\p{F}_A$, which can be evaluated through the work generated by each of the two (\emph{reactive} and \emph{active}) forces (see SI). The PRW displays a ratio $70/30\%$ between the components while the CPRW only has a $3\%$ \emph{reactive} contribution, which nevertheless is capitally important to reproduce spatial features of the walk. 

\section*{Discussion}

The present work explores the subject of human mobility in mid-range and non-crowded but mildly dense environments, from both empirical and theoretical perspectives by analysing and modelling the results of an experiment with pedestrians. The main contribution of the present paper consists in elucidating the minimal ingredients necessary to reproduce human mobility features in the context of a fair, where external and internal motion drivers are at play. 

Our non-phenomenological modelling approach proposes an additive, constructive and simple framework where pedestrians \emph{feel} the attraction of a collection of points of interest located in the fair. Starting with the most simple model (purely random walkers), we progressively introduce persistence and perceptual landscape (with attractive potential wells) aiming to untangle \emph{active} (inherent factors) and \emph{reactive} (external) contributions by only adding up to three parameters in the model (see Table~\ref{tab1}). The combination of persistence and perceptual landscape under a Langevin dynamic framework succeeds in explaining the most important aspects in this context: the non isotropic character of the human motion, the exponential decay of stopping times statistics due to the attraction of activities in stands, the exponential decay in flight lengths statistics during movement phases (finding a highly peaked velocity distribution). The framework is also able to quantify the relative importance of \emph{reactive} and \emph{active} components, showing that even a relatively small and subtle \emph{reactive} contribution can describe the basic traits of human motion in these contexts.

Directional memory effects are apparent in the way the different stands are visited, while a limit to the attention capacity of the participants in the fair is observed in the large-scale regime of about 10 minutes present in the stop duration statistics. The influence of the environment is patent from the spatial concentration around attention poles (attraction wells) and orientation of exploring patterns observed in the data. The presented framework is flexible enough to accept further enhancements. It would be possible, for instance, to add extra ingredients to account for destination selection or interaction with the crowd -- which might better explain the longest flights or the strong homogeneity in participants' velocities. We limit ourselves here to examining how the CPRW model might be extended to look at the destination selection effect, leaving other potential enhancements for future work. 

Our modelling approach contrasts that of Continuous Time Random Walks (CTRWs) \cite{weiss:1994}, because it aims to describe movement, which is an inherently time-continuous process, emerging from the interaction of several well-defined factors such as the presence of potential attractive wells \cite{Gardiner}. It then allows to test whether observed features from the data (such as Stop duration statistics) can be recovered under the hypotheses of the model, furthermore, it allows to explore the effects of data treatment and discretization and understand their biases. Last but not least, it further permits to quantitatively evaluate the balance between these components by means of physical magnitudes due to the straightforward interpretation one can extract from their computation.

Despite the success of the models in reproducing some of the observed features, the bimodal nature of these types of movement (stop-and-run states) together with human cognitive factors determining stand destination selection present a severe limitation to their description by means of a time continuous stochastic framework. Notwithstanding this, the presented models can be used to reproduce a variety of real world situations such as movement through parks, fairs, exhibition rooms and many other relatively small spaces with a medium-low density of individuals. The BP framework is flexible enough to easily modify the location of poles of attention (attraction wells) and then simulate the resulting pedestrian movements as people are attracted to them. This flexibility can be of great interest for re-designing spaces or anticipating pedestrian mobility in alternative spatial distributions.

Human mobility is a very complex process involving cognitive, physical and socio-economic factors, as well as a mix of spatial and temporal scales \cite{Brockmann2006}. Moreover, despite the efficiency of the different technologies used for tracking purposes, different technical limitations arise in each particular scenario, making its study an interesting cross-disciplinary area where technical, experimental and theoretical challenges from very different disciplines can meet. The present paper presents the results of one such cross-disciplinary collaboration, where modelling of human-related mobility in short ranges is covered. The junction of the open culture embedded in citizen science together with the endless possibilities in the advances in information technologies open new and interesting opportunities like the one exploited here. We hope that more experiments will follow and they will help tackle the challenges that the study of human mobility in medium ranges present.

\section*{Methods}

\subsection*{Experimental Setting.} 
During the weekend of 16 and 17 June 2012, the Institut de Cultura de Barcelona (Culture Institute of the Barcelona City Council) organised the Festa de la Ci\`encia i la Tecnologia (Science and Technology Fair). It is considered to be the most important annual science outreach event in Barcelona for general audiences. In 2012, it was organised as a set of fourteen stands and buildings located in an area of more than 4 hectares (330m $\times$ 145m) inside the public green park called Parc de La Ciutadella as shown in the top panel of Fig. \ref{fig1}. The event was held during the afternoon of Saturday from 16h to 24h and the morning of Sunday from 11h to 15h. Several activities took place within each of the stands, offered by researchers and science communicators. Their format was diverse, including games, experiments, astronomic observation, debates, micro-talks, workshops, and even performances. The participants had very different interests, origins, backgrounds and ages and the event organisers estimated that 10\,000 people visited the fair.
The Bee-Path information stand was located at one of the entrances of the Parc de la Ciutadella as shown in Fig. \ref{fig1} (spot number 1). This entrance was the most crowded access to the park. Visitors were encouraged to participate in the experiment by wandering around the fair while being tracked by an open source mobile app capable of running on a wide class of mobile phones. The raw data gathered is freely accessible from the project webpage: \url{www.bee-path.net}

\subsection*{Stop-and-run walks and rectangular grid criteria algorithms: Flights and Stops.}
GPS locations are classified in two mutually exclusive groups: stopped and moving points. We used a two step procedure inspired by Rhee et al. \cite{Rhee2007}. The algorithm firstly assigns by default a \emph{stopped} flag to a given point. If the next point in the time sequence is further away than a certain threshold distance $R_{stop}$, such a point is then flagged as \emph{moving}, otherwise the point remains as \emph{stopped}. Starting and ending points locations of individual tracks are always by default \emph{stopped} points. We chose a $R_{stop}=8$ m which is larger than the estimated GPS errors ($2$ m to $6$ m). SI justifies this choice. The second step of the procedure solely considers the locations flagged as \emph{moving} by grouping them using the rectangular grid criteria \cite{Rhee2007}: A flight for individual $u$ at time $t$ is defined as the minimal sequence of $N$ consecutive \emph{moving} locations $\{\p{x}_1,\p{x}_2...\p{x}_N\}$ that fit inside the box defined by the segment $\p{x}_N-\mathbf{x}_{1}$ and width corresponding to a parameter $R_{flight}$. Figure \ref{fig1} summarizes the two-step algorithm. Finally, despite the fact that the participants in the experiment had different demographic characteristics, they do not present highly distinctive population deviations in their movements (see SI) and we have thus aggregated their tracks in our study, considering them as statistically equivalent.

\subsection*{Dynamic Brownian Bridge Movement Model: the population-level utilisation density.}

We applied a dBBMM \cite{Kranstauber12,Bullard99,HorneEtal07} to each set of real, RW, PRW, CRW, CPRW, and Destination Selection Walk (DSW, see SI) model tracks (excluding those with less than 11 locations), using the results to generate a population-level utilisation density surface on a raster of 8 m cells placed over the study area. The dBBMM assigns probabilities based on the assumption of diffusion between observed locations. The model accounts for elapsed time as well as location error, and the dynamic version, proposed by Kranstauber et al \cite{Kranstauber12}, allows the diffusion coefficient to vary depending on the variance of the trajectory observed through a sliding window. We set the location error for all points at 4.071 meters, the mean of the GPS accuracy in the real data, and we used a sliding window of 11 points for the trajectory variance. We implemented the dBBMM calculations using the Move package by M. Smolla and B. Kranstauber (2015) for R (\url{http://www.R-project.org}). We also calculated the home range (95\% and 99\% isopleths) of each probability landscape, using the Geospatial Modelling Environment Version 0.7.2.1. Complementary GIS processes were done using ArcGis 10.1 (ESRI, ArcGis 10 on-line help \url{http://help.arcgis.com/en/arcgisdesktop/10.0/help/}). Maps were done using the © ortophotomap 1.5000 owned by the Catalan Cartographic Institute (ICC) and available at \url{www.icc.cat}.

\subsection*{Random component in the Langevin equation.} The stochasticity of Eq. (\ref{eq:langevin}) is represented in polar coordinates. The radial part is determined by a random variable $\rho$ which controls the modulus of the velocity and is distributed as $p(\rho) = \rho \exp({-\rho^2/2})$. The angular part is described by the random variable $\theta(t + \Delta t) = \theta(t) + \Theta$,
whose difference with the previous step follows the Von Mises Distribution
$p(\Theta) = \exp\left[ \frac{\kappa}{\Delta t} \cos (\Theta) \right]/[2\pi I_0\left( \frac{\kappa}{\Delta t}\right)],$
where $I_0$ is the modified Bessel function of order $0$ and $\kappa$ accounts for the persistence of the movement. When $\kappa=0$, we recover a RW, conversely, when $\kappa$ increases the probability density function becomes sharper around zero introducing reinforcement in the movement orientation and thus obtaining the CRW dynamics.

\subsection*{Bee-Path framework model simulations.}
\begin{figure}[htbp]
\includegraphics[width=1\textwidth]{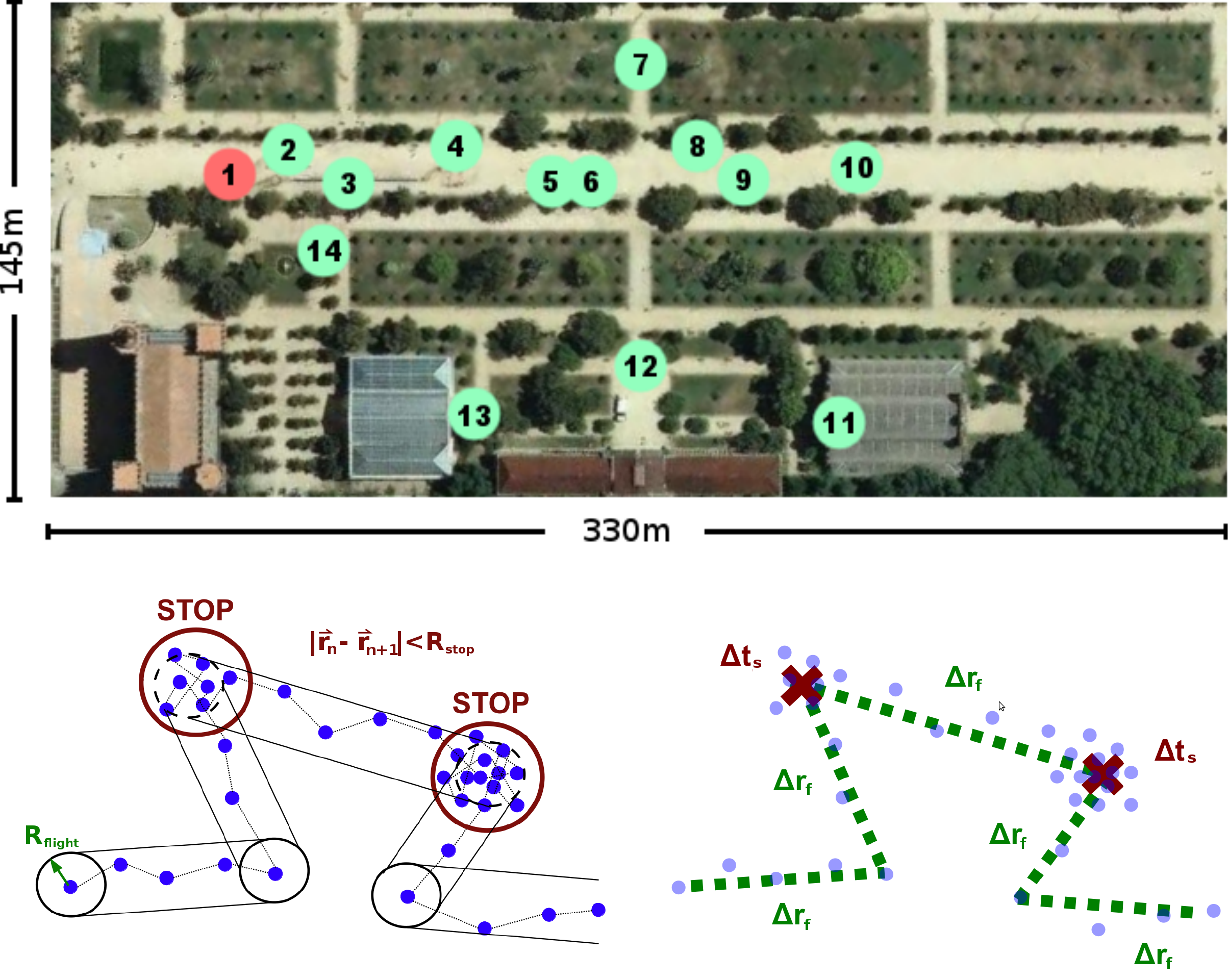}
\caption{Map of the fair indicating location of stands and stops and flights discrimination procedure. (top) Stand number one, coloured in red, shows the location of Bee-Path welcoming location, where participants were recruited. This image is produced from an orthophoto of ``Institut Cartogr\`afic i Geol\`ogic de Catalunya'' under a Creative Commons license CC-BY. (bottom) Each blue spot represents a GPS position recorded with a given time-stamp. Stop-and-run algorithm described in Methods is applied detecting the red circles of radius $R_{stop}$ as stopping points whose duration is $\Delta t_s$. Then, rectangular grid criteria whose width is calibrated with parameter $R_{flight}$ (bottom left) detects in this case five different flights with Euclidian lengths $\Delta r_f$ (bottom right).
}
\label{fig1}
\end{figure}

\begin{figure}[htbp]
\begin{center}
\includegraphics[width=1\textwidth]{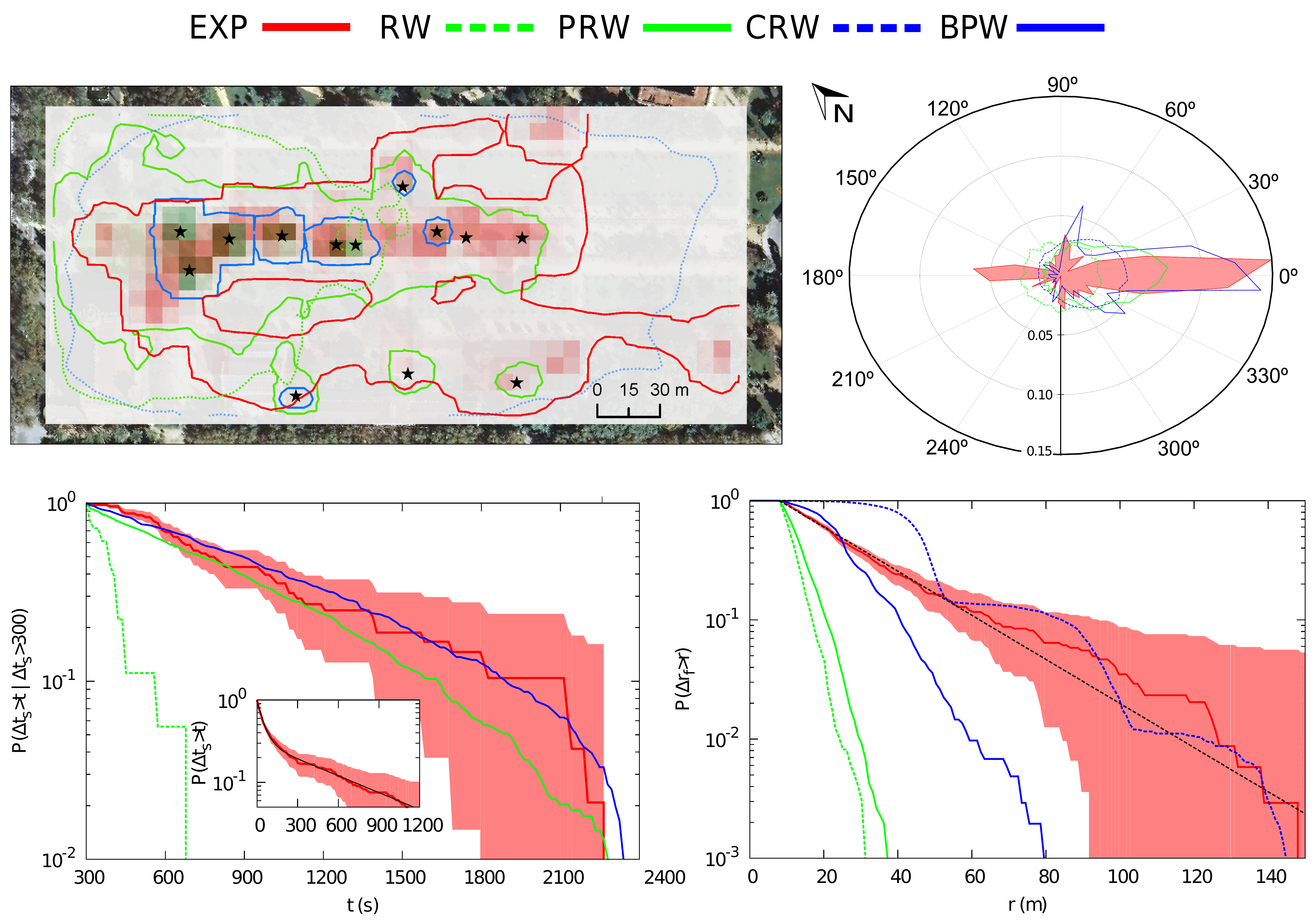}
\caption{Empirical results (EXP, red) compared with dynamics of each Bee-Path framework model: RW (green dashed), CRW (blue, dashed), PRW (green, solid), and CPRW (blue, solid). (top left) The dBBMM utilization densities, shown as coloured raster cells (2.5 standard deviation colour stretch, with darker colours indicating higher use intensities) and as $95\%$ isopleths, overlaid on ortophotomap of Ciutadella Park from ``Institut Cartogr\`afic i Geol\`ogic de Catalunya''. Attraction wells (stands) indicated by stars. (top right) Flight orientation polar plot where radial component measures the probability at the corresponding angle. RW, CRW, PRW, CPRW jointly with the results from the empirical flights. (bottom left) The stop duration Complementary Distribution Function (CDF) for long stops ($\Delta t_s > 300s$) is calculated for each dynamics with a bin size of $\Delta t_{s}=15$ seconds. Shaded area around the real curve represents the cumulative standard deviation calculated as $\sqrt{\sum_{0}^{i}p_{j}(1-p_j)/N}$, where $p_j$ is the value of bin $j$ and $N$ is the number of stops. The inset shows the complete CDF Stop duration distribution for the real time and the analytical function (black solid line) with $\omega = 0.308\pm 0.008$, $\tau_1 = 617\pm17$s and $\tau_2 = 37.5\pm 1.5$s ($\chi^2_{red} = 6.57\times 10^{-5}$). (bottom right) Flight length CDF is calculated and dashed black line correspond to heuristic fit with $\lambda=27\pm2$m ($\chi^2_{red} = 0.008$). The CRW shows a bumpy behaviour because of the combination of time discretization process (every 15 seconds) and directional memory of the model.}
\label{fig2} 
\end{center}
\end{figure}

\begin{figure}
\includegraphics[width=1\textwidth]{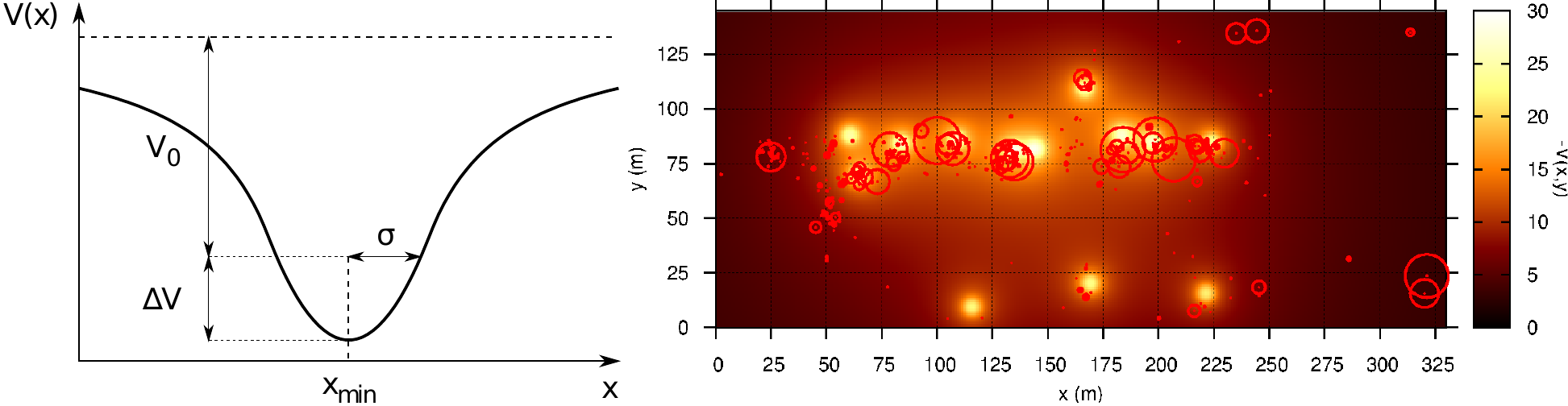} 
\caption{Proposed gravitational wells representing attractiveness of fair stands. (left) Section across x-axis direction of one potential well. (right) Heat map of fair's potential landscape, with potential wells placed at the fourteen stands locations. All participants' stops indicated by red circles with radius proportional to stop duration. Note that longest stops closely coincide with stand locations.}
\label{fig3}
\end{figure}

The four proposed scenarios simulate $20\,000$ seconds ($5.5$ hours) of real experiment. Simulations take place inside an area of $330$ m by $145$ m displayed in Fig. \ref{fig1}. Attraction wells are placed in the same configuration as the stands in the fair, creating a potential-well landscape. In every realization $50$ individuals are simultaneously created at the same spot where participants downloaded the GPS application in the fair (spot 1, see Fig. \ref{fig1}), each of them with the same lifetime as a real subject picked at random. When an individual has reached his lifetime, has visited all the wells or has crossed the borders, it is removed and another one is created following the same procedure. Simulation time-step $\Delta t$ is set at $0.1$ seconds but individuals' locations are sampled every $15$ seconds and fed into the aggregation algorithm for their analysis, mimicking the data gathering process of the real experiment.

\paragraph{Supplementary Information} is linked to the online version of the paper at www.nature.com/nature.

\paragraph{Acknowledgements.} 
We greatfully acknowledge the participation of the anonymous citizen scientist volunteers who
made this research possible. We are indebted to Barcelona Lab programme through the Citizen Science Office promoted by the Direction of Creativity and Innovation from the Barcelona City Council led by I Garriga for their
help and support for setting up the experiment in Parc de la Ciutadella and for giving us the opportunity to include the experiment in the Barcelona Science Fair. We are also indebted to Mar Canet for developing the Bee Path app and many other programs; and to David Rold\'an, Marc Angl\'es, and Berta Paco for providing support during the experiment. We also want to thank Ionas Erb, Jos\'e Espinosa-Carrasco, Carole Faviez, C\'edric Notredame, and Mara Dierssen for helpful discussion during the preparation of the experiment. The research leading to these results has received funding by Barcelona City Council (Spain), RecerCaixa (Spain) through grant {\it Citizen Science: Research and Education} (FB, JRP, AO, MGR and JP), by Strep-EU LASAGNE Project under contract No. 318132 (ADG and OS), by MINECO (Spain) through grants FIS2013-47532-C3-2-P (JP and MGR) and FIS2012-38266-C2-2 (ADG and OS), by Generalitat de Catalunya (Spain) through grants 2014-SGR-608 (MGR, OS, JP and OS) and 2012-ACDC-00066 (MGR, OS, JP and OS), and by Fundaci\'on Espanyola para la Ciencia y la Tecnolog\'ia (FECYT, Spain) through the Barcelona Citizen Science Office project of the Barcelona Lab programme (JP). OS also acknowledges financial support from Generalitat de Catalunya (FI-program) and the Spanish MINECO (FPU-program).

\paragraph{Author contributions:} 
M.G.-R., O.S., A.D.-G., J.P. and F.B.\ designed research,
M.G.-R., O.S., A.D.-G., J.P., A.O, J.R.B.P. and F.B. \ performed research and wrote the paper;
M.G.-R., O.S. and A.O. \ analyzed the data, and M.G.-R. and O.S.\ proposed and studied the theoretical model.
All authors read and approved the final version.


\paragraph{Competing financial interests:} 
The authors declare no competing financial interests.

\begin{table}[htbp]
\begin{center}
\begin{tabular}{c|cc|cccc|cc}
Movement & \multicolumn{2}{c}{Dynamics} & \multicolumn{4}{c}{Observables Reproduced} & \multicolumn{2}{c}{Components}\\ 
& Potential & Persistence & Util. Dens. & Flight Orient. & Stops & Flights & \emph{Active} & \emph{Reactive}\\ \hline
RW & NO & NO & - & - & - & - & $100\%$ & $0\%$\\
PRW & YES & NO & + & + & ++ & - & $32\%$ & $68\%$\\
CRW & NO & YES & - & - & - & + & $100\%$ & $0\%$ \\
CPRW & YES & YES & + & ++ & + & + & $97\%$ & $3\%$\\
\end{tabular}
\caption{Overview of characteristics and agreement with empirical data for Random Walk (RW), Potential driven Random Walk (PRW), Correlated Random Walk (CRW) and Correlated Potential driven Random Walk (CPRW) models. ``Dynamics'' columns indicate whether dynamics are driven by a potential and whether they incorporate persistence. ``Observables'' columns compare each model's dynamics with empirical utilization denisty, flight orientation, and stop and flight Complementary Distribution Functions (cf. Fig. \ref{fig2}), using (-) to indicate disagreement, (+) moderate agreement, and (++) good agreement. ``Components'' columns measure the importance of \textit{reactive} and \textit{active} components in each case (see SI for further details). Note that PRW shows a slight tendency to flight orientation, but less intense than CPRW, while CPRW is able to reproduce flight distribution when probability of skipping wells is introduced in the model as discussed in SI. As also shown in SI, velocity distribution of CPRW is wider and larger (median and average) than that empirically observed while PRW better coincides with observations.}
\label{tab1}
\end{center}
\end{table}

\end{document}